# The Utility of the Virtual Imaging Trials Methodology for Objective Characterization of AI Systems and Training Data

Fakrul Islam Tushar[1,2], Lavsen Dahal[1,2], Saman Sotoudeh-Paima[1,2], Ehsan Abadi[1,2], William P. Segars[1,2], Joseph Y. Lo[1,2], Ehsan Samei[1,2]

[1]Center for Virtual Imaging Trials, Carl E. Ravin Advanced Imaging Laboratories, Dept. of Radiology, Duke University School of Medicine, Durham, NC 27705, USA

[2]Department of Electrical & Computer Engineering, Pratt School of Engineering, Duke University, Durham, NC 27705, USA

Corresponding author: Fakrul Islam Tushar (e-mail: tushar.ece@duke.edu).

## ABSTRACT

**Purpose:** The credibility of Artificial Intelligence (AI) models for medical imaging continues to be a challenge, affected by the diversity of models, the data used to train the models, and applicability of their combination to produce reproducible results for new data. In this work, we aimed to explore whether emerging Virtual Imaging Trials (VIT) methodologies can provide an objective resource to approach this challenge.

**Approach:** The study was conducted for the case example of COVID-19 diagnosis using clinical and virtual computed tomography (CT) and chest radiography (CXR) processed with convolutional neural networks. Multiple AI models were developed and tested using 3D ResNet-like and 2D EfficientNetv2 architectures across diverse datasets.

**Results:** Model performance was evaluated using the area under the curve (AUC) and the DeLong method for AUC confidence intervals. The models trained on the most diverse datasets showed the highest external testing performance, with AUC values ranging from 0.73-0.76 for CT and 0.70-0.73 for CXR. Internal testing yielded higher AUC values (0.77-0.85 for CT and 0.77-1.0 for CXR), highlighting a substantial drop in performance during external validation, which underscores the importance of diverse and comprehensive training and testing data. Most notably, the VIT approach provided objective assessment of the utility of diverse models and datasets, while offering insight into the influence of dataset characteristics, patient factors, and imaging physics on AI efficacy.

**Conclusions:** The VIT approach enhances model transparency and reliability, offering nuanced insights into the factors driving AI performance and bridging the gap between experimental and clinical settings.

# I. INTRODUCTION

Radiology artificial intelligence (AI) models often struggle to generalize, resulting in limited clinical applicability [1]. This is primarily due to the existential limits on the diversity of data upon which a model can be trained. Obviously, models trained on small datasets, such as from a single center, do not represent the general population. But even large, multi-center datasets can still be plagued by variability across scanner vendors, acquisition protocols, reconstruction algorithms, pre/post-processing, or patient inclusion criteria. Consequently, AI developers continuously strive for massive amounts of data, hoping that the large magnitude can overcome the generalizability problem.

Failure of medical imaging AI models to generalize is a pervasive problem. The crisis of reproducibility was starkly evident during the COVID-19 pandemic when chest radiography (CXR) and computed tomography (CT) were initially employed for detecting and managing lung infections [2, 3]. In the rush to develop AI aides for radiologists, however, many studies reported unrealistic, near-perfect performances that dropped almost to chance upon external testing [4-10]. The availability of numerous large public datasets of medical images [11-15], including those from the Medical Imaging and Data Resource Center (MIDRC) [16], has led to a plethora of AI models for the diagnosis of COVID-19. Yet a review of 62 studies asserted that none of these models were fit for clinical use due to methodological flaws and underlying biases [17]. While the focus on COVID-19 is waning and imaging is no longer used for the primary diagnosis of the disease, with some exception [18], the rare combination of so much data accompanied by widespread problems (e.g., limited documentation, heterogeneous acquisition standards, and inconsistent labeling) in reproducibility offers our field a rare opportunity to understand how best we can appropriate AI methods, for both clinical practice as well as future health crises. For example, commonly used saliency methods such as Gradient-weighted Class Activation Mapping (Grad-CAM) have been reported to suffer from reproducibility issues [19], underscoring the need for more objective frameworks.

A promising solution to this ongoing reproducibility crisis in AI lies in the use of the Virtual Imaging Trial (VIT) approach. In this work, we define reproducibility as the objective and repeatable evaluation of an AI system under controlled and well-characterized variations in patient anatomy, imaging systems, and acquisition protocols, such that consistent performance trends are observed when experiments are repeated. Simulating the three key components of an imaging trial, patients, scanners, and readers [20], VITs offer control over input variables to generate virtual images representing a diverse range of patient characteristics and imaging techniques. Diseases can be simulated with pixel-level ground truth in terms of their location, size, and characteristic features. Simulations can also encompass different scanner technologies or acquisition protocols. Given the precise controls used to generate this data, we hypothesize that VITs can elucidate which factors drive model performance. This approach has the potential to facilitate not only truly independent external validation but also rigorous and unbiased testing across diverse scenarios.

The VIT approach has been applied to a wide range of diseases and modalities, including lesion detection in mammography and breast tomosynthesis [21-23], nodule detection, and COPD quantification in chest CT. Previous work involved VITs for validating deep-learning models for COVID-19 detection using clinical and virtual datasets [24-26].

This study aimed to ascertain how the VIT methodology can add objectivity, explainability, and overall generalizability to the AI process. The study was done in the context of COVID-19 diagnosis, given its associated diversity of AI models and data. Building upon our prior validated, open-source deep-learning models for case-level COVID-19 detection with CT and CXR images [24, 25], the study modeled virtual patients replicating a diverse range of anatomies and manifestations of COVID-19 pneumonia [27, 28]. The virtual patients were imaged using simulated scanners, replicating the physical and technical characteristics of actual medical imaging devices. The AI model “readers,” representing expert-readers, read the images, allowing us to evaluate the diagnostic performance of AI models under consistent conditions. Comparing with results from multiple clinical datasets and multiple AI models, we aimed to unpack the interplay of dataset-model matching and mismatching on the results, to evaluate the influence of patient- and physics-based factors on the generalizability of the results, compare CT and CXR modalities, and to assess the utility of VIT as an independent validation to provide a controlled environment for evaluating AI models.

## II. METHODS

Institutional Review Board approval was obtained for this exempt study. The study, detailed below, deployed anonymized clinical image datasets as well as simulated virtual data. The clinical datasets were multiple [11-15, 29-33], varying in size, diversity, demographics, and class definitions. The virtual data were from a population of 4D-XCAT models with varying COVID-19 size and distribution, imaged using virtual CT and CXR scanners (DukeSim, CVIT, Duke University) [27].

Multiple convolutional neural network (CNN) models (detailed below) with residual connections were developed to process CT and CXR images efficiently. These lightweight CNNs, designed to reduce computational complexity while maintaining high accuracy, were used to classify cases as positive or negative for COVID-19. The CNN models were trained using single and various combinations of clinical datasets. In parallel experiments, CT and CXR clinical datasets were used to evaluate model performance on internal versus external validation, allowing assessment of potential performance shifts.

The virtual data were reserved as a separate external validation dataset. In addition, by varying the parameters of the virtual imaging trials, we systematically assessed how model performance was affected by factors pertaining to the patients (i.e., infection size) or imaging physics (i.e., effective dose and modalities). An illustration of the overall workflow of the analysis is presented in Figure 1.

### II.A. Clinical Dataset

The clinical CT data used in this study included a total of 12,844 volumes from 7,452 patients across 10 datasets: RICORD [33], MosMed [11] BIMCV-COVID-19 +/- (BIMCV-V2) [12], COVID-CT-MD [13], CT Images in COVID-19 [34], PleThora [32], COVID19-CT-dataset [15], Stony Brook University COVID-19 Positive Cases (COVID-19-NY-SBU) [14], A Large-Scale CT and PET/CT Dataset for Lung Cancer Diagnosis (Lung-PET-CT-Dx) [29], and Lung Image Database Consortium / Image Database Resource Initiative (LIDC-IDRI) [35]. These datasets had different prevalences of COVID-19 positive and negative images (Figure 2a) and demographics. Summary statistics of the CT datasets are detailed in Table 1.

Furthermore, all ten clinical CT datasets above were combined to create the U-10 CT dataset, which provides a more diverse dataset for factors such as patient population and demographics, disease appearances, CT systems, and imaging protocols. We applied pre-cleaning procedures: removing non-axial and non-chest protocols, reformatted or partial scans, studies with fewer than 20 slices, and cases with undefined or inconsistent COVID-19 test results. Figure 4 shows the inclusion and exclusion criteria followed in the curation of the clinical CT data. These cleaning steps were necessary for consistency but were not excessively time intensive.

**TABLE 1.** Clinical CT patient datasets utilized in model development and testing. The combination of all ten constitutes the U-10 CT dataset. Demographic values are reported as the percentage of patient sex and mean of patient age.

| No | Dataset | Source | Demographics | Category | Train* | Validation* | Test* |
|---|---|---|---|---|---|---|---|
| **1.** | RICORD [33] (1b,1b) | Turkey, USA, Canada, Brazil | 44% women | COVID+ | 66 (90) | 22 (32) | 22 (33) |
| | | | Age 54 ±17 | COVID- | 70 (72) | 23 (23) | 24 (25) |
| | | | | **Total** | **136 (162)** | **45 (55)** | **46 (58)** |
| **2.** | MosMed [11] | Russia | 56% women Age 47 | COVID+ | 512 (512) | 170 (170) | 174(174) |
| | | | | COVID- | 152 (152) | 50 (50) | 52 (52) |
| | | | | **Total** | **664 (664)** | **220 (220)** | **226 (226)** |
| **3.** | BIMCV-V2 [12] | Spain | 42% women. | COVID+ | 455 (1421) | 152(484) | 152(470) |
| | | | Age 64 ±16 | COVID- | 728 (2077) | 239(706) | 268(823) |
| | | | | **Total** | **1183 (3498)** | **391 (1190)** | **420 (129)** |
| **4.** | COVID-CT-MD [13] | Iran | 40% women. | COVID+ | 101(101) | 33 (33) | 35 (35) |
| | | | Age 51 ±16 | COVID- | 81 (81) | 27 (27) | 28 (28) |
| | | | | **Total** | **182 (182)** | **60 (60)** | **63 (63)** |
| **5.** | An et al. [34] | Multi-center | N/A | **COVID+** | 379 (391) | 126 (129) | 127 (130) |
| **6.** | PleThora[32] | USA | 31% women. Age 68 ± 10 | **COVID-** | 241 (241) | 80 (80) | 81 (81) |
| **7.** | COVID19-CT [15] | Iran | 39.1% women Age: 47 ± 16 | **COVID+** | 604 (604) | 201 (201) | 202 (202) |
| **8.** | COVID-19-NY-SBU[14] | USA | 43% women. (Age: ranges between 18-90 years) | **COVID+** | 251 (739) | 84 (278) | 84 (282) |
| **9.** | Lungs-CT-Dx [29] | China | 46% women, Age 61 ± 10 | **COVID-** | 207 (479) | 69 (154) | 70 (164) |
| **10.** | LIDC-IDRI [35] | USA | N/A | **COVID-** | 606 (611) | 202 (204) | 202 (203) |
| | **Total / U-10 CT** | | | | **4453 (7571)** | **1478 (2571)** | **1521 (2702)** |

**Note**-* Number of patients (number of scans), COVID+= COVID-19 positive, COVID-= COVID-19 negative, COVID-19-NY-SBU = Stony Brook University COVID-19 Positive Cases, Lungs-CT-Dx= A Large-Scale CT and PET/CT Dataset for Lung Cancer Diagnosis.

CXR analysis included 25,219 clinical CXR images collected from three datasets: Fricks et al. [31], BIMCV [12], and COVIDx-CXR-2 [30]. These datasets also had different prevalences of COVID-19 positive and negative images (Figure 2b) and demographics. All three clinical CXR datasets were also combined to form the U-3 CXR dataset, with one important caveat. In one of the datasets, COVIDx-CXR-2, positive images were from different sources, but the negative class was much larger and mainly from one source, namely the RSNA Pneumonia Detection Challenge [36] (Figure 2b). To ensure a balanced training and validation process for the unified U-3 dataset, the negative cases were randomly subsampled to achieve an equal distribution between the two classes, which also provided a more balanced contribution from this particular dataset. Summary statistics of the CXR datasets are detailed in Table 2.

**TABLE 2.** Clinical CXR Patient Cohorts utilized in model development and testing. Demographic values are reported as the percentage of patient sex and mean of patient age.

| No | Dataset | Source | Demographics | Category | Train | Validation | Test |
|---|---|---|---|---|---|---|---|
| **1.** | Fricks et al.[31] | Iran, Italy, USA | N/A | COVID+ | 544 | 136 | 171 |
| | | | | COVID- | 174 | 44 | 55 |
| | | | | **Total** | **718** | **180** | **226** |
| **2.** | BIMCV-V2 [12] | Spain | 46% Women<br>Age 63 ± 17 | COVID+ | 2694 | 674 | 843 |
| | | | | COVID- | 2265 | 566 | 708 |
| | | | | **Total** | **4959** | **1240** | **1551** |
| **3.** | COVIDx-CXR-2 [30] | Multi-center | N/A | COVID+ | 1727 | 431 | 200 |
| | | | | COVID- | 11034 | 2759 | 200 |
| | | | | **Total** | **12761** | **3190** | **400** |
| | **Total** | | N/A | | **18438** | **4610** | **2177** |
| 4 | **U-3 CXR dataset** | | N/A | COVID+ | 4965 | 1241 | 1214 |
| | | | | COVID- | 4965 | 1241 | 963 |
| | | | | **Total** | **9930** | **2482** | **2177** |

## II.B. Virtual Dataset

The XCAT computational phantoms used in this study were based on the method described in detail by Abadi et al. [27]. An overview of the method is illustrated in Figure 3. Creating computational phantoms for COVID-19 is a process that unfolds in four distinct stages: constructing the body framework, detailing the morphological characteristics of lung abnormalities, replicating the texture and composition of affected lung tissues, and performing simulated scans to generate the virtual images.

**Body Framework Construction:** The process began with the development of the normal anatomy using the 4D extended cardiac-torso (XCAT) model from Duke University [37, 38]. The XCAT model provided a comprehensive foundation with detailed anatomy, dynamic organ motions, and textured tissues, built from real patient data spanning a range of patient characteristics such as sex, body size, and lung volume. Fifty separate phantoms were used for this study.

**Insertion of COVID-19 Abnormalities:** The second stage involved the meticulous detailing of lung abnormalities typical of COVID-19, such as ground-glass opacities (GGO) and consolidations. This was achieved by examining CT scans from clinically confirmed COVID-19 patients (N=20), where the abnormalities were manually segmented and modeled in a series of surfaces mimicking the morphology [27]. Next, the lung parenchyma's texture was adjusted to reflect the changes observed in actual CT images, such as the addition of fluids in the case of GGO or the uniform texture seen in consolidations.These adjustments followed the COVID-19 XCAT framework described by Abadi et al. [27], in which the material composition and texture of abnormal regions were tuned so that the mean HU and contrast-to-noise in regions of interest matched representative patient CT scans. These modeled features were then integrated into the XCAT phantoms, ensuring a match in body dimensions, sex, and age, to represent the disease's manifestations within the computational models accurately.

**Simulating the Imaging Process:** Virtual CT and CXR datasets were generated by scanning the virtual patient models with or without the disease using an X-ray image acquisition simulator (DukeSim, CVIT, Duke University) [27, 28]. DukeSim is designed to replicate the physical processes in x-ray imaging with CT and CXR, including modeling of x-ray tube spectra, scanner geometry, and detector configuration. DukeSim combines ray tracing for rapid image generation and Monte Carlo techniques for accurate modeling of scatter, attenuation, and detector noise. The virtual framework allowed the scanning of the same virtual patient with both modalities without other confounding factors. Virtual scans were repeated at different effective doses (0.01, 0.1, 0.3, 1.6, 5.6, and 11.2 mSv). The dose settings were selected to represent a wide range of clinical applicability, as well as a direct comparison of CT and CXR images at the same hypothetical dose and motion state. For the CXR acquisitions, two commercial post-processing algorithms (denoted as algorithms A and B to maintain confidentiality) were applied to examine the effects of vendor heterogeneity. Table 3 shows the characteristics of the generated CT and CXR images. Using a parallel computing cluster with eight 48 GB GPUs, we achieved an average of 36 virtual CT scans per hour and generated each CXR scan in under one minute.

**TABLE 3.** Virtual (CVIT-COVID) dataset attributes, including imaging protocols and disease distributions.

| | Number of virtual exams | |
|---|---|---|
| **Effective dose (mSv)** | **COVID-19** | **Negative** |
| | **CVIT-COVID-CT** | |
| **0.3** | 50 | 40 |
| **1.6** | 50 | 40 |
| **5.6** | 50 | 40 |
| **11.2** | 50 | 40 |
| **Total (CT)** | **200** | **160** |
| | **CVIT-COVID-CXR** | |
| **0.01** | 50 | 40 |
| **0.10** | 50 | 40 |
| **0.3** | 50 | 40 |
| **Total (CXR)** | **150** | **120** |

## II.C. Pre-Processing

Standard preprocessing was performed on both CT and CXR images [39, 40]. Each CT volume was resampled to voxel dimensions of 2 mm × 2 mm × 5 mm (w, h, d). Intensities were clipped between -1000 to 500 HU, then standardized to a mean of 0 and standard deviation of 1. To reduce computational cost and the influence of background organs, three-dimensional (3D) patches of size 160×160×96 (w, h, d) were centered about the lungs. The patch size was based on average lung size plus a margin to allow for patient variability. CXR images were resized and randomly cropped to a size of 300x384 pixels, then standardized to 0.5 mean and 0.5 standard deviation to maintain consistency across both the train and test datasets.

## II.D. Model Development and Training

As noted above, previous studies have shown that complex deep learning models can produce non-generalizable near-perfect performance due to fundamental overtraining [24, 25]. To minimize this effect, we intentionally selected lightweight ResNet-like models [40, 41] and trained four separate CT-based models using the RICORD, MosMed, BIMCV, and U-10 CT datasets. The ResNet architecture has also shown consistent performance across various medical imaging tasks [24, 40, 41]. Similarly, for CXR, we trained four different EfficientNetv2 [42] models using the data from Fricks et al., BIMCV, COVIDx-CXR-2, and U-3 CXR datasets, respectively. Each dataset was randomly divided by the patient into subsets of training (60%), validation (20%), and testing (20%). No cross-validation was performed; instead, we utilized a train-validation-test split. As we aimed to assess the utility of virtual data for clinically trained algorithms, no training was applied to the virtual data. Instead, the model trained on clinical data was applied to the entire virtual dataset for testing, which served exclusively as a test set.

CT AI-models used a simple 3D CNN inspired by ResNet [43], the architecture is shown in Figure 5. After initial convolution, features were learned across two resolution scales, then halved by max-pooling (pooling size 2×2×2) while doubling the number of filters. The last R-block features underwent batch normalization, rectified linear unit (ReLu), global max-pooling, dropout (dropout rate 0.5), and finally, a dense classification layer with sigmoid activation for binary case-level COVID-19 detection. Additionally, we applied L2 regularization with a coefficient of 0.001 to prevent overfitting. The stochastic gradient descent (SGD) optimizer was used to optimize the weights with decay learning rate, and weighted binary cross-entropy was used as the loss function. Weights were initialized to a uniform distribution. To retain the natural prevalence, no class balancing was performed during training.

The hyperparameters for the CT AI-models were set as follows: initial learning rate of 1e-6, maximum learning rate of 1e-4, learning rate decay of 1e-2, batch size of 24, and 300 training epochs. CXR models were based on Efficientnetv2 with the original architecture [42], SGD was selected as the optimizer with the learning rate scheduler [44], initial learning rate of 0.01, and cross-entropy loss. All models were developed using Python TensorFlow v2.6 and PyTorch deep learning frameworks. Model training utilized lightweight architectures, requiring ~8–16 GB VRAM to balance efficiency and performance. All model weights, initial hyperparameters, and code are made publicly accessible [45].

## II.E. Evaluation and Statistical Analysis

We conducted a series of evaluations to assess the model performance on clinical and virtual datasets. For each clinical dataset, we followed established procedures from prior studies by performing patient-level binary classification for the presence or absence of COVID-19 [5, 10, 24, 30]. For the virtual dataset, the controlled simulation process allowed us to evaluate further the influence of input variables related to the patient as well as the image acquisition.

We first evaluated the effect of the patient-related factor of infection size to understand the impact of infection severity on model performance. In our simulated XCAT phantoms with embedded COVID-19 abnormalities, the infection volume was quantified as a percentage of total lung volume. The median infection percentage across cases (2.6%) was then used as the threshold to divide virtual COVID-19 pneumonia cases into two groups: 'higher' infection (≥2.6%) and 'lower' infection (<2.6%). This approach helps in assessing how well the AI models perform across a spectrum of disease severity and identifying any performance biases or limitations. Additionally, we conducted evaluations based on the physics of image acquisition, specifically the imaging modality of CXR vs. CT, as well as a wide range of effective radiation doses.

To support our findings and assess the significance of the results, all performances were evaluated using the receiver operating characteristic area under the curve (AUC) with 95% confidence interval (CI) calculated by the DeLong method as implemented by pROC 1.16.2 in R 3.6.1 with 2000 bootstrapping samples [46].

## III. RESULTS

### III.A. Evaluation of Models' Performance on Clinical Data

As depicted in Figure 6, clinical CT and CXR models exhibited a consistent drop in performance from internal to external testing, and those differences often exceeded the confidence intervals. While some loss of performance is expected in external testing, these remarkably consistent differences indicate systemic differences across these datasets. The CT AI-models showed an internal validation AUC range of 0.69 to 0.85, whereas external testing consistently dropped to between 0.54 and 0.76. Similarly, for CXR models, internal performance ranged from an AUC of 0.77 to 1, while external testing AUC again dropped to a range of 0.51 to 0.73. Models trained on the most diverse datasets (U-3 CXR and U-10 CT) consistently yielded a testing performance that was the highest or second highest. Notably, despite its size, the COVIDx-CXR-2 dataset for CXR exhibited substantial dataset bias, with positive and negative cases drawn from different sources and a large class imbalance (Figure 2b), which likely contributed to the near-perfect internal performance of the COVID-19 CXR classification models and optimistic external testing results, even for the U-3 model trained on all three datasets. To facilitate inspection of operating points, full ROC curves for all models are provided in Supplementary Figs. S1–S2. The modest AUC values (often <0.7) indicate that these networks are not suitable for direct clinical use and should be interpreted only in this comparative context

### III.B. Evaluation of Models' Performance on Virtual Data

As shown in Figure 6, all CT AI-models achieved intermediate AUC values on the virtual data, consistent with their performance on the combined clinical training data. In this sense, the virtual data produced performance that was comparable to, and in some cases not worse than, that observed on several clinical external test sets, suggesting that the simulations are sufficiently realistic to support AI characterization in this setting.

Among the CT AI models, the model trained on the most diverse U-10 CT dataset achieved the highest AUC on the virtual CT images, but overlapping confidence intervals indicate it is not definitively better than the other CT models. A similar pattern was observed with the CXR models, further supporting the usability of the virtual datasets.

### III.C. Evaluation per Patient and Disease

Assessing the effect of infection extent, Figure 7 shows all models performed better on both CT and CXR images for higher infection compared to images with lower infection. For the lower infection groups, performance was near chance (AUC ~0.5) for all CXR models and most CT models. Although CXR performance was generally limited, the most highly infected cases may be detectable. These results are consistent with expectations and suggest that differences in disease severity mix across datasets may account for performance variability. This study also demonstrates the utility of VIT towards clinical explainability.

### III.D. Evaluations per Image Acquisition

For the same virtual patients, we assessed the performance of models over a wide, overlapping range of effective doses for the virtual CT and CXR acquisitions. As shown in Figure 8, the 3D CT AI-models consistently outperformed the 2D CXR models, but the confidence intervals for the AUCs overlapped. Within each modality, although the effective dose (mSv) varied by 30-fold to represent the widest possible range of clinical use, there was no statistically significant change in performance [38, 47]. These results are consistent with the task of patient-level classification of this diffuse disease where quantum noise has limited impact, whereas other tasks involving small or high-frequency structures (e.g., lung nodules) may exhibit stronger dose dependence.

## IV. DISCUSSION

There has been considerable work on AI models to improve radiology diagnosis. However, the practical application of these models in clinical practice has been hindered by two related challenges. First, models often underperform when applied to a new dataset with different attributes. Second, most models function as “black boxes” that lack interpretability, making it difficult to determine which factors may contribute to an outcome or poor performance. These issues became particularly evident during the urgent scientific response to COVID-19, when many early studies reported high performances that did not generalize [24, 25, 48-51]. Although biases in AI models for healthcare may be unavoidable, where we use ‘bias’ to mean a systematic mismatch between the training data distribution and the intended clinical population (e.g., site imbalance, label noise, or case enrichment), a comprehensive understanding of such factors, supported by effective external testing, can raise confidence that such models are trustworthy [17, 49, 52]. This study addresses the problem of biases in medical imaging AI models by leveraging clinical and virtual data for independent testing, thus enabling the evaluation of both generalizability and interpretability. While the study used COVID-19 diagnosis as its target task, given its rich resources of models and data, the lessons learned can apply to other AI-targeted tasks.

We compiled a large cohort of clinical CT and CXR images from dataset resources representing over 22,000 patients. Despite the large amount of training data, however, model performance was still impaired due to class imbalance and confounding issues such as radiographic markers, incorrect image orientation, and collimator edges [49, 50, 53]. Proper data curation is time-consuming and requires domain expertise in medical imaging, rendering this process prohibitively costly [39]. Therefore, external validation of AI models is essential to rule out biases [17, 49]. Towards that goal, this study explored the use of virtual data [20, 27], which proved to offer two crucial advantages. First, the virtual image data provided external validation that is not only truly independent but also controlled. Second, the VIT framework allowed the evaluation of the models under different patient- and physics-based factors. This provided an opportunity for interpretability with respect to clinical or technical dependencies. Our study demonstrates the utility of VIT simulations to conduct AI imaging studies in a trustworthy, reproducible, and practicable manner.

One of our primary objectives was to analyze the impact of dataset variability on model generalization. To minimize overfitting, we intentionally used lightweight networks [40, 41]. Even so, all models still dropped in performance substantially from internal to external testing, consistent with other studies [24, 25, 49, 50]. This generalizability gap likely reflects inherent biases in the existing datasets with regard to institutional bias, patient demographics, disease appearances, and image quality [49, 51, 53]. To address such bias, we trained models on the combined U-10 CT and U-3 CXR datasets, which incorporated greater diversity. These models demonstrated improved external testing performance compared to the single-dataset models. The model trained on the diverse U-10 CT dataset demonstrated a very consistent AUC of approximately 0.73 across all three clinical datasets, suggesting that combining diverse data yields more credible and representative performance for this challenging clinical task. These general trends were also observed for the CXR datasets, but with considerable residual bias due to the disproportionate influence of the COVIDx-CXR-2 dataset, which is much larger than other datasets and leads to confounding bias as its positive and negative cases come from different institutions. This quandary shows that despite rigorous training and external testing, AI models can still be affected by fundamental data biases.

The VIT process proved to deliver a controlled portrayal of true clinical performance. When many models were tested on virtual images, they performed consistently within the range of external testing on clinical datasets, suggesting that the simulations presented data that may support AI characterization in this setting.Unlike clinical datasets, the virtual images are free from institutional bias or other confounding factors, because the VIT framework offers precisely reproducible controls in terms of patient sampling as well as physical image formation. This enabled us to compare identical virtual patients with and without the disease (in 40 of the 50 cases, COVID-19 abnormalities were embedded into the same XCAT model) and also to conduct virtual imaging of each patient using both CT and CXR. The degree of experimental control provided by VITs is not physically possible in real clinical trials.

By integrating virtual datasets with clinical datasets, we aimed to enhance the generalizability and reliability of AI systems in medical imaging, ensuring their applicability in diverse clinical scenarios. The concept of a virtual dataset is integral to our study, providing a robust alternative to conventional datasets. These datasets offer precise control over imaging parameters, including patient anatomy, disease characteristics, and imaging conditions, ensuring consistency and reproducibility. Unlike conventional datasets, which often suffer from variability in patient demographics and imaging protocols, virtual datasets enable a controlled and repeatable generation of imaging data. As shown in Table 4, virtual datasets possess advanced features such as comprehensive patient-level, slice-level, and pixel-level annotations, and the ability to image the same virtual patient with both CT and CXR at multiple doses. These features facilitate rigorous evaluation and validation of AI models, allowing for systematic studies of the effects of various factors on model performance.

**TABLE 4.** Attributes of CT and CXR datasets. Note that virtual data are the only ones that contain all attributes, including the advanced features where the same virtual patient can be imaged with both CT and CXR at multiple doses, with multiple CXR post-processing. X= available.

| Datasets | Class Type | | Label Level | | | Advanced features |
|---|---|---|---|---|---|---|
| | COVID-19 positive | COVID-19 negative | Patient-level | Slice-level | Pixel-level | |
| **CT datasets** | | | | | | |
| RICORD[33] | X | X | X | | | |
| MosMed[11] | X | X | X | | | |
| BIMCV-Iteration 2[12] | X | X | X | | | |
| COVID-CT-MD[13] | X | X | X | X | | |
| An et al. dataset[34] | X | | X | | | |
| COVID19-CT-dataset[15] | X | | X | | | |
| COVID-19-NY-SBU[14] | X | | X | | | |
| Lungs-CT-Dx[29] | | X | X | | | |
| LIDC-IDRI[35] | | X | X | | | |
| Duke-CVIT-CT | X | X | X | X | X | X |
| **CXR datasets** | | | | | | |
| Fricks *et al*. dataset[31] | X | X | X | N/A | | |
| BIMCV-2[12] | X | X | X | N/A | | |
| COVIDx-CXR-2[30] | X | X | X | N/A | | |
| Duke-CVIT-CXR | X | X | X | N/A | X | X |

Our VIT analysis further provided intriguing insights into the effects of patient- and physics-based factors driving AI performance. Regardless of the training datasets for both the CT and CXR models, there was a noticeable increase in performance when the COVID-19 infection size was larger than the median value. For both imaging modalities, performances stayed consistent even across a 30-fold range in effective dose (which well exceeds the range in clinical practice) [54, 55], suggesting that dose may not be as relevant for the AI detection of diffuse diseases such as pneumonia. In stark contrast to the model evaluation on clinical data, our analysis confirmed that CT outperformed CXR, which was consistent with expectations since 3D CT scans provide superior spatial information over 2D CXR images [56, 57]. Prior studies have similarly reported higher sensitivity of CT for COVID-19 detection [57, 58].

This study had several limitations. Although the virtual CT and CXR images realistically reproduced both anatomical and physical processes, they were generated from a pool of fifty virtual patients with variable anatomy and severity of the disease. By using large, diverse datasets and testing on independent virtual data, this study yielded results that were more reliable than prior studies. Nonetheless, the lower performances confirm that these models are not ready for clinical deployment and are used here only as testbeds for comparative evaluation. Virtual datasets offer control and reproducibility but must be complemented with real-world validation to ensure ethical and clinically applicable AI models. In our study, simulation testing showed consistent trends but with large confidence intervals. Future work will increase the number of computational phantoms to represent even larger and more diverse patient populations and explore the inclusion of additional imaging parameters to improve realism. Models were developed only to conduct case-level detection, which is the only annotation available in almost all datasets. Furthermore, the label of COVID-19 as negative or positive was defined independently per each dataset, and those standards varied widely, including radiologist assessment or different diagnostic tests [2]. Some datasets included both COVID-19 pneumonia and other types of pneumonia, which may not be readily differentiated by imaging alone. Finally, future work should aim to address these limitations by incorporating more detailed multi-class annotations and evaluating model performance under different disease classification scenarios.

## V. CONCLUSIONS

AI-based diagnosis models hold the potential to revolutionize healthcare. However, factors contributing to model bias remain underexplored, especially in the medical imaging domain. An essential prerequisite to clinical deployment is a robust external evaluation. The VIT framework plays a crucial role in addressing the ongoing reproducibility crisis in AI models by providing the necessary image data that is objective and controlled. By enabling consistent evaluation across diverse scenarios such as different dose levels, disease extents, and imaging modalities (CT and CXR), VIT not only helps to identify bias but also facilitates improvements in model robustness and generalizability. As emerging AI techniques continue to evolve [59-61], the need for rigorous evaluation frameworks like VIT becomes even more critical to ensure their reliability and clinical relevance. By studying patient- or physics-based factors influencing model performance, the VIT methodology offers potential for interpretability and opportunities for model refinement. Through these contributions, virtual imaging trials can complement clinical trials and evaluations by providing controlled pre-testing that streamlines trial design and makes them faster, more rigorous, and more reproducible.

## DISCLOSURES

The authors have no conflicts of interest to declare that are relevant to the content of this article.

## APPENDIX A – DATA AVAILABILITY

The clinical data utilized in this study are open-source and can be referenced via the citation in Table 1 and Table 2. The authors are committed to promoting transparency and open science. Reasonable requests for access to an anonymized version of the private datasets (Duke-CVIT-CT and Duke-CVIT-CXR) can be made by contacting the corresponding author. Upon publication, all model weights, initial hyperparameters, and code will be publicly accessible at https://gitlab.oit.duke.edu/cvit-public/cvit_revicovid19

## ACKNOWLEDGMENT

This work was funded in part by the Center for Virtual Imaging Trials, NIH/NIBIB P41EB028744 and NIH/NCI R01 CA261457.

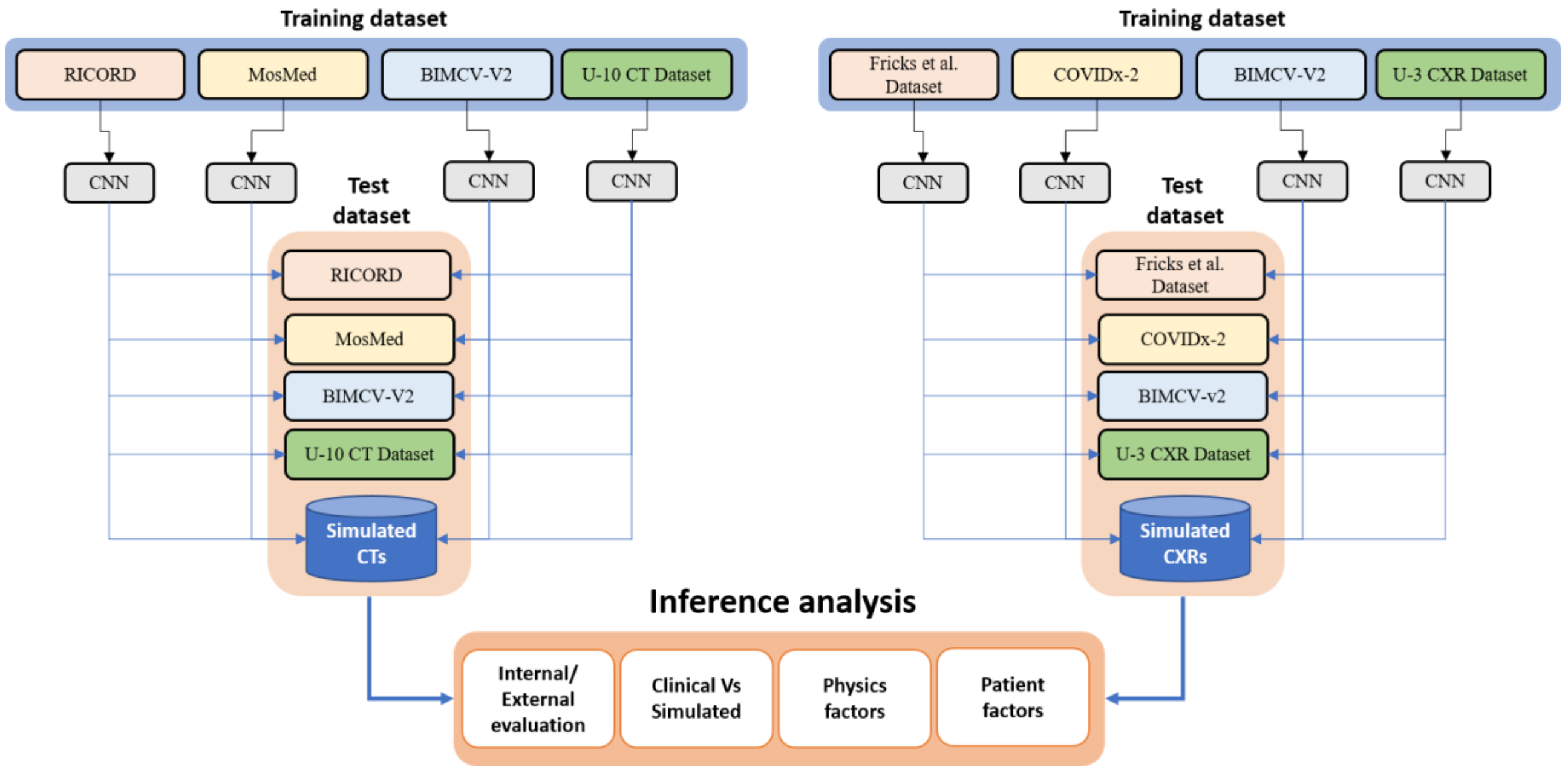

**FIGURE 1.** Study design overview. 12,844 CT scans and 25,219 CXR images for COVID-19 diagnosis were drawn from 13 clinical datasets comprising single or multiple centers (Supplement Fig. 1-2). Multiple deep-learning-based models were developed using these clinical datasets. All models underwent internal testing (held-out from the same training dataset) and external testing (all other datasets). Further external testing was performed using virtually simulated CT and CXR images to analyze effect of patient and imaging physics factors.

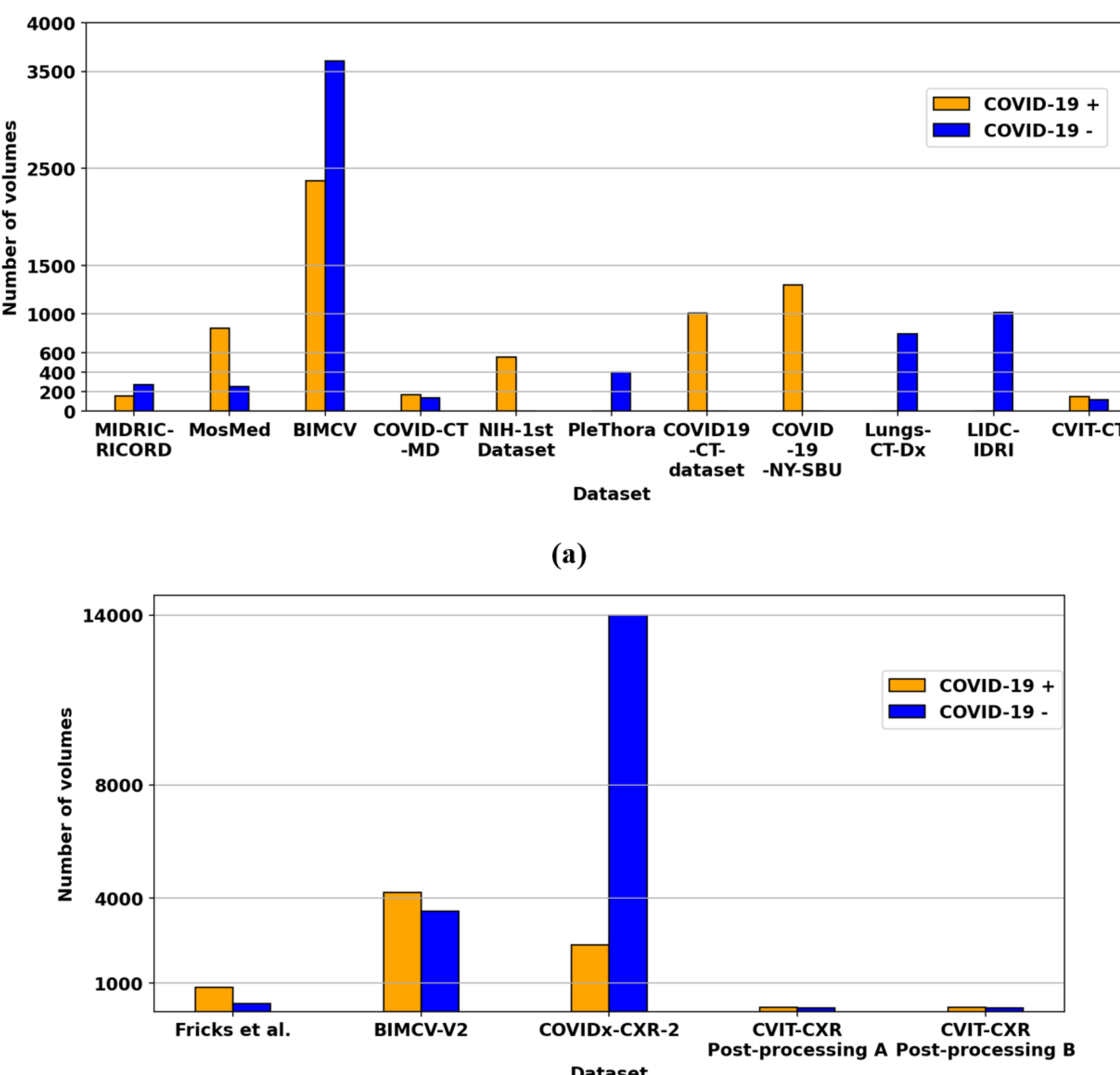

**FIGURE 2.** Histograms showing distribution of COVID-19 positive (+) and negative (-) cases among different datasets (clinical and simulated) (a) CTs and (b) CXRs. Log scale is used to show the large variation in numbers of exams. Note that the prevalence varies greatly, and some datasets contain only one class.

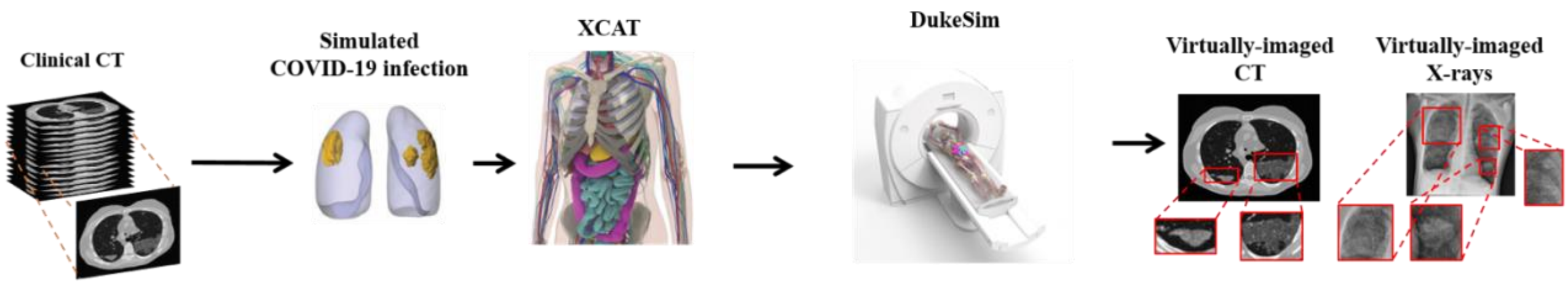

**FIGURE 3.** An overview COVID-19 computation phantoms development and simulated CT and CXR images.

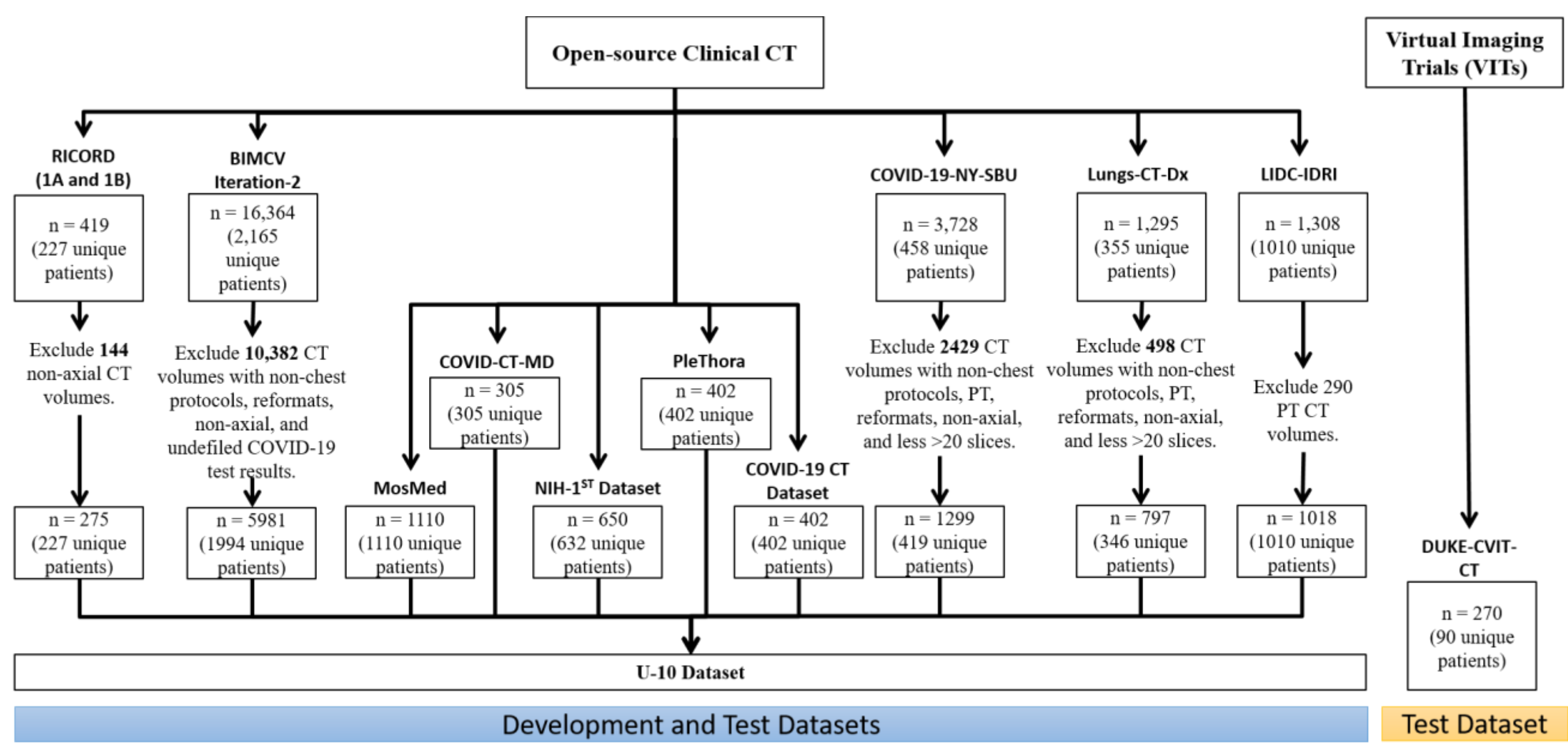

**Figure 4.** Flowchart of inclusion and exclusion criteria for the chest CT scans. n= number of CT volumes. A total of 16,949 CT scans of 11,166 patients were used for model development and testing. There were ten clinical datasets: RICORD [33], MosMed [11], BIMCV-COVID-19 +/- (BIMCV-V2) [12], COVID-CT-MD [13], CT Images in COVID-19 [34], PleThora [32], COVID19-CT-dataset [15], Stony Brook University COVID-19 Positive Cases (COVID-19-NY-SBU) [14], A Large-Scale CT and PET/CT Dataset for Lung Cancer Diagnosis (Lungs-CT-Dx) [29], and Lung Image Database Consortium / Image Database Resource Initiative (LIDC-IDRI) [35], These ten clinical datasets were united into the U-10 CT Dataset. Additionally, simulated data were from the Center for Virtual Imaging Trials CT Dataset, Duke-CVIT-CT [27].

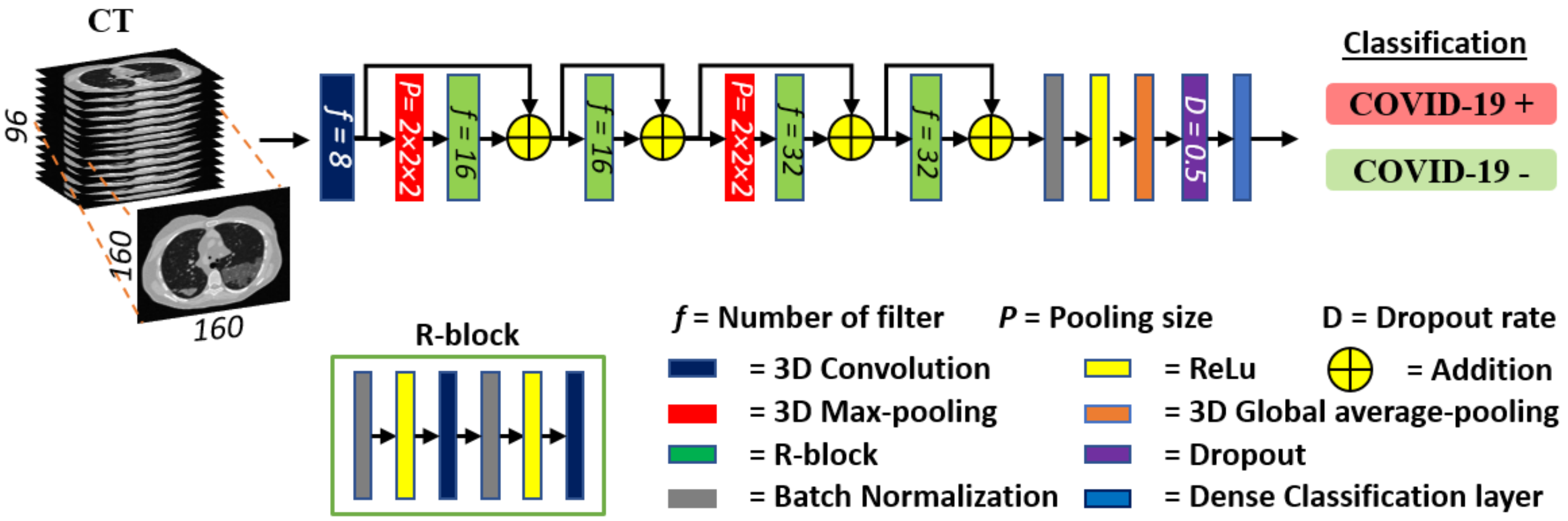

**Figure 5.** 3D CNN architecture for CT classification of COVID-19. The classification module is a 3D Resnet-like model with 2 R-Blocks in each resolution. The number of filters is denoted as f. The final output is a tensor of the probability of being COVID-19 positive or negative.

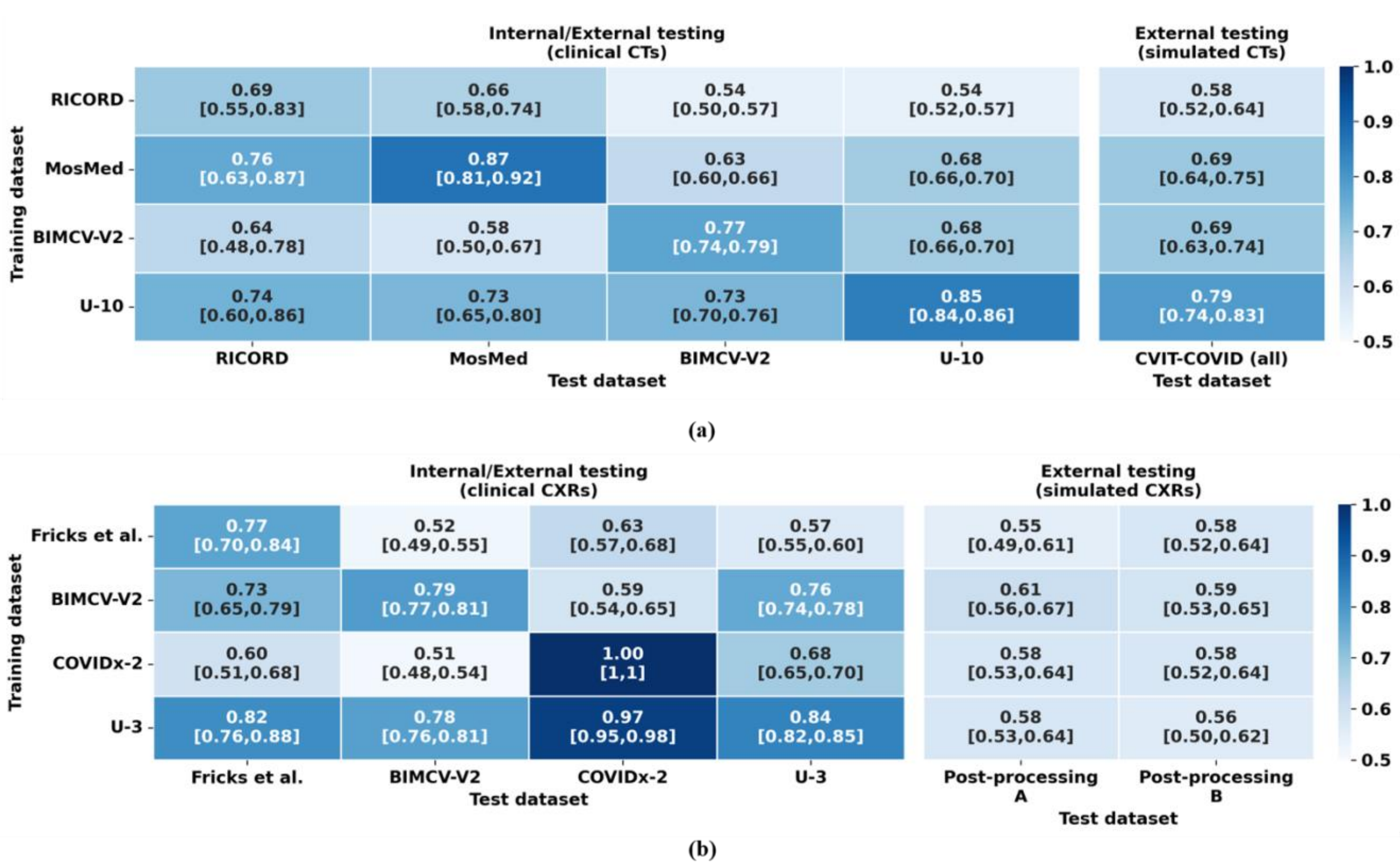

**Figure 6.** Confusion matrix of case-level COVID-19 detection performance of (a) CT and (b) CXR models. Training dataset is shown in rows and testing dataset in columns; diagonal represents internal validation, while off-diagonal entries are external testing. Additional external testing on simulated images is shown on the right. Performance is reported as receiver operating characteristic area under the curve with 95% confidence interval.

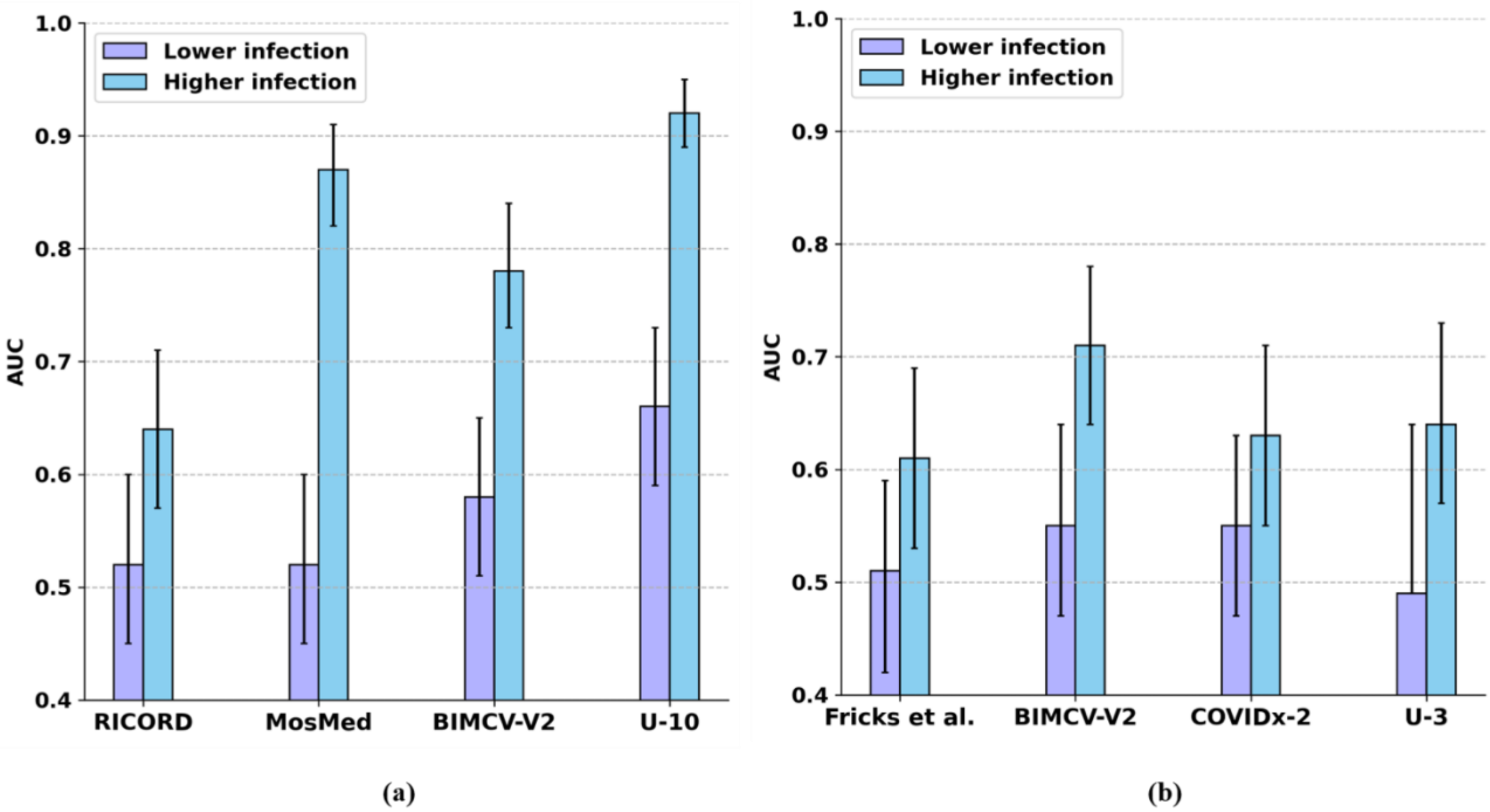

**Figure 7.** Both (a) CT and (b) CXR models each trained on four datasets (represented on the x-axis), consistently demonstrated superior performance in "higher infection" cases, where the pneumonia volume exceeded the median, compared to "lower infection" cases that fell below the median. For CXR, results were almost identical for the two post-processing algorithms, so only algorithm A is shown. Error bars represent the 95% confidence interval.

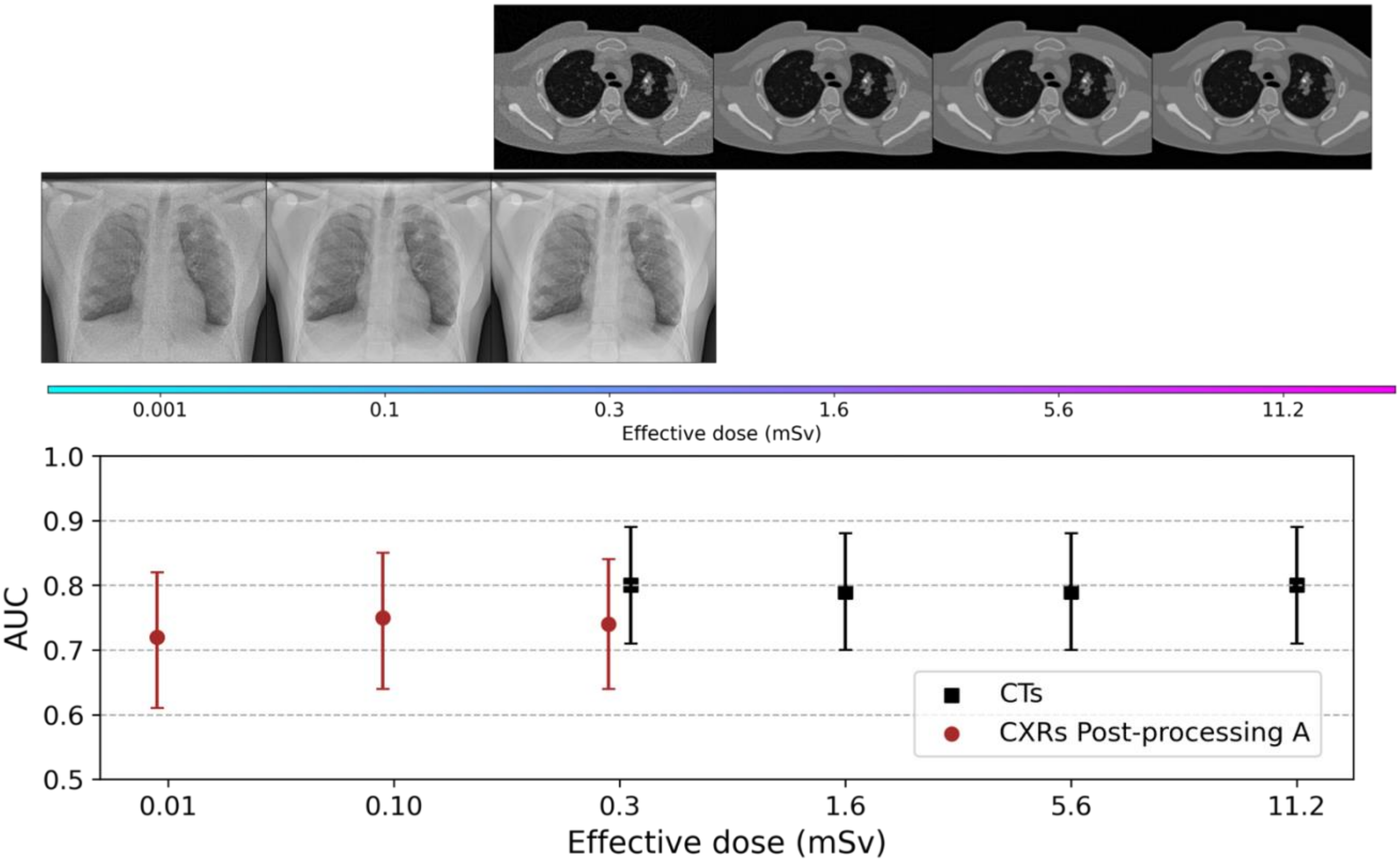

**Figure 8.** Simulated images were used to evaluate physics-based factors. Although models consistently performed better on CT over CXR, the differences were not significant at the shared dose of 0.3 mSv. Within each modality, performances were also not significantly different across a wide range of effective dose. Error bars correspond to 95% confidence interval.

# Supplements

Tushar F.I, et al. The Utility of the Virtual Imaging Trials Methodology for Objective Characterization of AI Systems and Training Data.

**Figure S1.** CT Testing Receiver operating characteristic (ROC) curves for models trained on four clinical datasets: (a) RICODRD, (b) MosMed, (c) BIMCV-V2, and (d) U-10 and tested on clinical and simulated datasets. In each panel, the model is trained on the training split of the dataset named in the panel and evaluated on its corresponding test datasets.

**Figure S2.** CXR Testing Receiver operating characteristic (ROC) curves for models trained on four clinical datasets: (a) Fricks et al., (b) BIMCV-V2, (c) COVIDx-CXR-2, and (d) U-3 and tested on clinical and simulated datasets. In each panel, the model is trained on the training split of the dataset named in the panel and evaluated on its corresponding test datasets.

**Figure S3.** Example of the two post-processing strategies of the simulated CXRs: (a) Post-processing A and (b) Post-processing B.

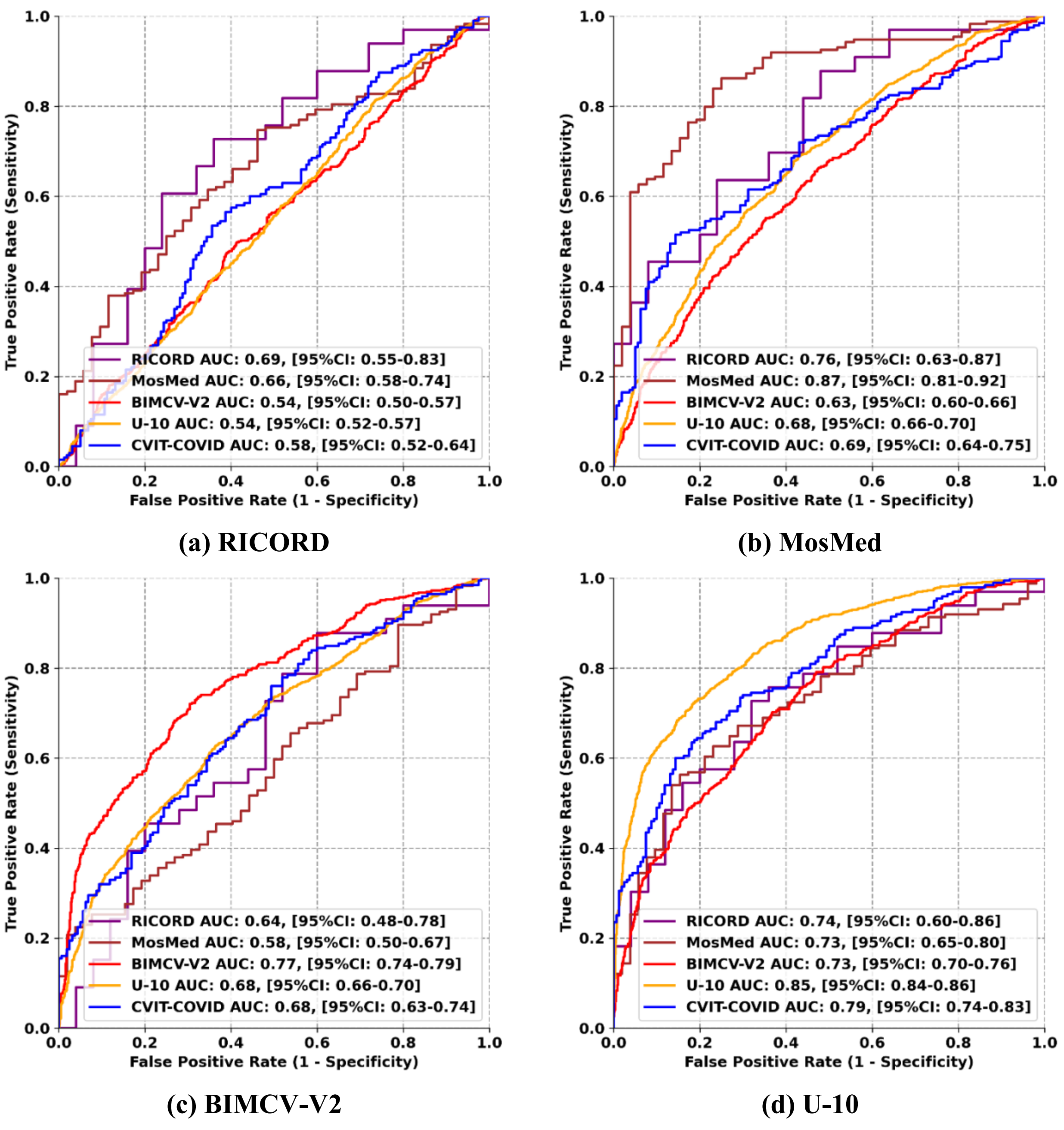

**Figure S1.** CT Testing Receiver operating characteristic (ROC) curves for models trained on four clinical datasets: (a) RICODRD, (b) MosMed, (c) BIMCV-V2, and (d) U-10 and tested on clinical and simulated datasets. In each panel, the model is trained on the training split of the dataset named in the panel and evaluated on its corresponding test datasets.

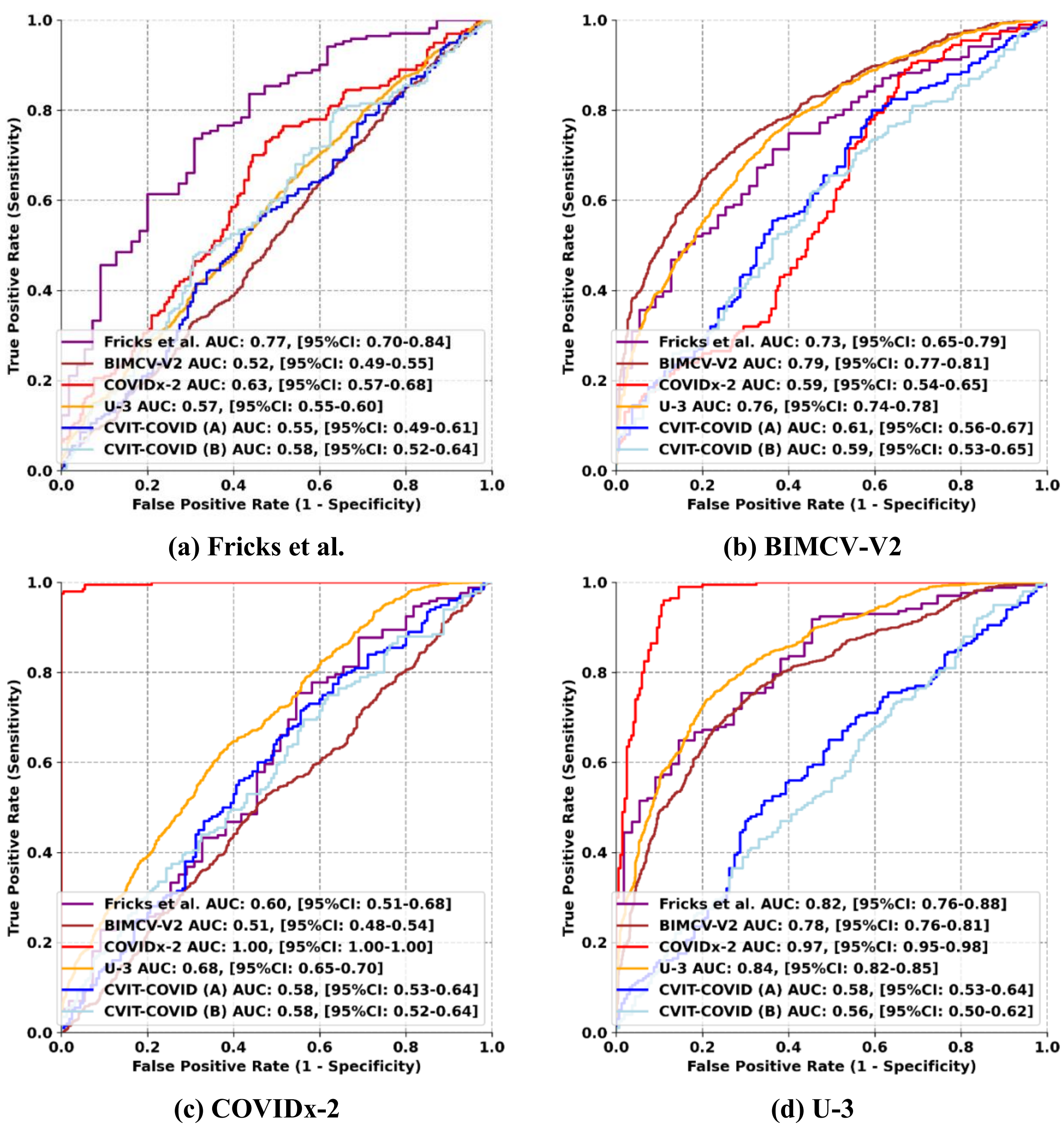

**(a) Fricks et al.** **(b) BIMCV-V2**

**(c) COVIDx-2** **(d) U-3**

**Figure S2.** CXR Testing Receiver operating characteristic (ROC) curves for models trained on four clinical datasets: (a) Fricks et al., (b) BIMCV-V2, (c) COVIDx-CXR-2, and (d) U-3 and tested on clinical and simulated datasets. In each panel, the model is trained on the training split of the dataset named in the panel and evaluated on its corresponding test datasets.

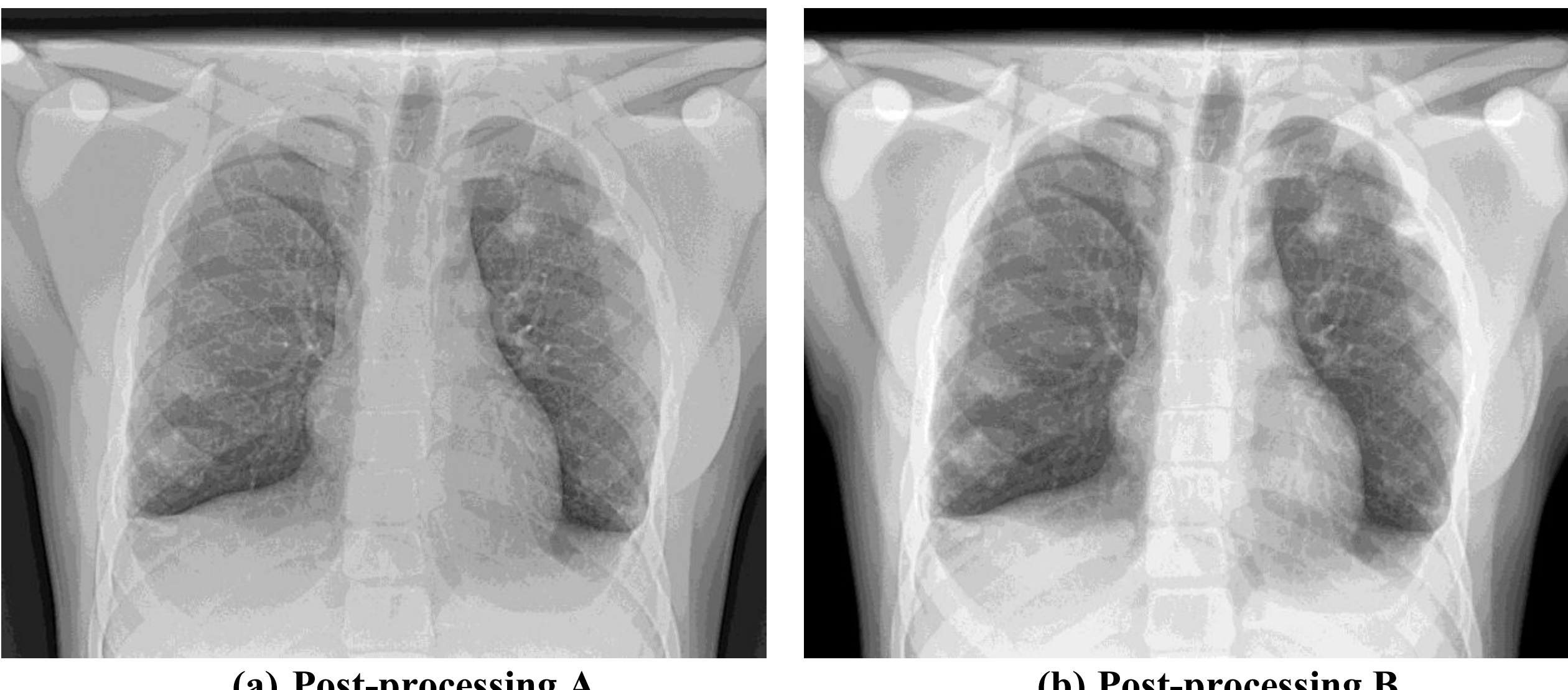

**(a) Post-processing A** **(b) Post-processing B**

**Figure S3.** Example of the two post-processing strategies of the simulated CXRs: (a) Post-processing A and (b) Post-processing B.